\documentclass[11pt]{article}

\usepackage{amsmath,amsfonts,amsthm,amssymb}

\newcommand{\eq}{\begin{equation}}
\newcommand{\en}{\end{equation}}
\newcommand{\eqa}{\begin{eqnarray}}
\newcommand{\ena}{\end{eqnarray}}

\begin{document}

 \setlength{\unitlength}{1mm}

 \vspace*{0.1cm}


\begin{center}

   {\bf Integrable Quantum Computation}
 \vspace{.5cm}

 Yong Zhang\footnote{
 Institute of Physics, Chinese Academy of Sciences, Beijing 100190, P.R. China}\footnote{zhangyo2008@gmail.com}

\end{center}
\begin{center}

\parbox{13.cm}{

 { \bf Abstract }
 Integrable quantum computation is defined as quantum computing via the integrable condition,
 in which two-qubit gates are either nontrivial unitary solutions of the Yang--Baxter equation
 or the Swap gate (permutation). To make the definition clear, in this  article,
 we explore the physics underlying the quantum circuit model, and then present a unified
 description on both quantum computing via the Bethe ansatz and quantum computing via the
 Yang--Baxter equation.

\vspace{1.cm}



 }
\end{center}


\section{Physics underlying the quantum circuit model}

Computers are physical objects, and computations are physical processes, which may be one of Deutsch's
famous quotes \cite{Deutsch89}. This statement gives rise to a natural question: what is the physics
underlying the quantum circuit model? Deutsch's answer  may be that quantum computing
votes for the many-worlds interpretation of quantum mechanics. Nielsen and Chuang's answer to it stays the same
in the two editions of their popular textbook \cite{NC2011}: ``outside the scope of the present discussion".
In our current understanding \cite{Zhang11},  the quantum circuit model can be viewed as a generalization
of the multi-qubit factorisation scattering model in integrable systems \cite{Sutherland04}, and the scalable
quantum computer is thus supposed to be  an exactly solvable model satisfying the integrable
condition \cite{Sutherland04}.

It becomes well known in quantum information science after DiVincenzo's work \cite{DiVincenzo95} that
an arbitrary $N$-qubit quantum gate can be expressed as a sequence of products of some two-qubit gates.
Hence a quantum circuit is described as a network of two-qubit gates, which may be called the
{\em locality principle} of the quantum circuit, see Preskill's online lecture notes \cite{Preskill97}.
On the other hand, it has been widely accepted for a long time in integrable systems \cite{Sutherland04}
that many-body factorisable scattering can be expressed as a sequence of two-body scattering.  As two-qubit
quantum gates are considered as two-qubit scattering operators, the quantum circuit is thought of
as {\em the generalized factorisable scattering model} defined as a generalization of the factorisation
scattering model \cite{Sutherland04}.

Sutherland in his book \cite{Sutherland04} describes three types of the integrable conditions which are capable
of yielding exactly solvable quantum many-body systems: the first is the Bethe ansatz; the second is the
Yang--Baxter equation; and the third is the quantum Lax equation. In physics, hence, we define {\em integrable
quantum computation} as {\em quantum computing via the integrable condition} so that we can obtain a unified
description on both quantum computing via the Bethe ansatz \cite{Zhang11} and quantum computing via the
Yang--Baxter equation \cite{ZKG05}. Note that we do not deal with quantum computing via the quantum Lax equation
in this paper.

Section 3 is a preliminary introduction on quantum computing via the Bethe ansatz  \cite{Zhang11}. The main
reference is Gu and Yang's paper \cite{GY89} on the application of the Bethe ansatz to the $N$ Fermion problem
with delta-functional potential. Given a quantum many-qubit system, i.e., the Hamiltonian is known, if the Bethe
ansatz is satisfied, then it is an integrable model  to yield many-qubit factorisable scattering as
a sequence products of two-qubit scattering matrices in which two-qubit gates satisfy the Yang--Baxter equation.

Section 4 is a brief sketch on quantum computing via the Yang--Baxter equation \cite{ZKG05}. Given a nontrivial
unitary solution of the Yang--Baxter equation, the Hamiltonian of an integrable system can be constructed
in principle. Additionally, the paper \cite{ZKG05} is  written in the way to emphasize how to derive unitary
solutions of the Yang--Baxter equation from unitary representations of the braid group. In view of the
research \cite{Zhang11}, however, it has been distinct that {\em integrable quantum computation} is a research
subject independent  of quantum computing via unitary braid representations.  Hence this section is a refinement
of the research \cite{ZKG05} from the viewpoint of {\em integrable quantum computation}.

In mathematics,  {\em integrable quantum computation} specifies a type of quantum circuit model of computation in
which two-qubit gates are either the Swap gate (permutation) or nontrivial unitary solutions of the Yang--Baxter
equation.

\section{Quantum computing via the Bethe ansatz}

Let us consider the model of $N$ qubits (spin-1/2 particles) in one dimension interacting via the delta-function
potential, with the Hamiltonian given by
 \eq
 \label{delta}
 H=-\sum_{i=1}^N \frac {\partial^2} {\partial x_i^2} + 2 c \sum_{1\le i <j \le N} \delta(x_i-x_j)
 \en
where $c>0$ ($c<0$) means repulsive (attraction) interaction, see  \cite{GY89} for the detail. Interested readers
are also invited to refer to Bose and Korepin's article \cite{BK11} for the physical reason why this model is
interesting in experimental quantum information and computation.

For two qubits at $x_1$ and $x_2$,  the outgoing wave in the region $x_1 < x_2$
with respective momenta  $k_1 < k_2$ is given by
\eq
 \label{two-out}
 \alpha_{out} \,  e^{i (k_1 x_1 + k_2 x_2) }, \quad  \alpha_{out}=\alpha_{12},
\en
the incoming wave in the region $x_1 < x_2$ is given by
\eq
\alpha_{21} \,  e^{i (k_2 x_1 + k_1 x_2) },    \quad  \alpha_{inc}=\alpha_{21},
\en
so that the two-qubit gate (i.e., the scattering matrix) $\check{R}_{12}(k_2,k_1)$ defined by $\alpha_{out}
 =\check{R}_{12} \alpha_{inc}$ has the form
  \eq
 \check{R}_{12}(k_2, k_1) =\frac {i (k_2-k_1)(-P_{12}) +c } {i(k_2-k_1) -c },
 \en
where the permutation operator $P_{12}$ exchanges the spins of two flying qubits.

For three  qubits at $x_1$, $x_2$ and $x_3$,  the outgoing wave in the region $x_1 < x_2 <x_3$
with respective momenta  $k_1 < k_2 <k_3$ is given by
\eq
\label{three-out}
 \alpha_{out} \,  e^{i (k_1 x_1 + k_2 x_2 +k_3 x_3) }, \quad  \alpha_{out}=\alpha_{123},
\en
the incoming wave in the region $x_1 < x_2 <x_3$ is given by
\eq
\alpha_{321} \,  e^{i (k_3 x_1 + k_2 x_2 + k_1 x_1) }, \quad  \alpha_{inc}=\alpha_{321},
\en
so that $\alpha_{out}$ is determined by $\alpha_{inc}$ in the  way
\eq
 \alpha_{out} = \check{R}_{12}(k_2,k_1)\,\check{R}_{23}(k_3,k_1)\, \check{R}_{12}(k_3,k_2)\,\alpha_{inc},
\en
or in the other way
\eq
\alpha_{out} = \check{R}_{23}(k_3,k_2)\,\check{R}_{23}(k_3,k_1)\, \check{R}_{23}(k_2,k_1)\,\alpha_{inc},
\en
which give rises to the consistency condition (the Yang--Baxter equation) on the three-qubit scattering,
\eq
\label{YBE-three}
\check{R}_{12}(u)\,\check{R}_{23}(u+v)\, \check{R}_{12}(v)\, = \,\check{R}_{23}(v)\, \check{R}_{12}(u+v)\,
  \check{R}_{23}(u)
\en
with $u=k_1-k_2$, $v=k_2-k_3$ and $u+v=k_1-k_3$.

The outgoing wave of two qubits at $x_1$ and $x_2$ in the region $x_1 < x_2$ with respective momenta
$k_1 < k_2$ is given by (\ref{two-out}), but the incoming wave in the region $x_2 < x_1$ is given by
\eq
 \alpha_{inc} \,  e^{i (k_1 x_1 + k_2 x_2) }, \quad  \alpha_{inc}=(-P_{12})\,\alpha_{21},
\en
so that the scattering matrix of two qubits in different regions has the form
\eq
 \alpha_{out} =\check{R}_{12}(k_2,k_1) (-P_{12})\alpha_{inc}.
\en
Similarly,  the outgoing wave of three qubits at $x_1$, $x_2$ and $x_3$ in the region $x_1 < x_2 <x_3$
with respective momenta  $k_1 < k_2 <k_3$ is given by (\ref{three-out}), but the incoming wave in the
region $x_3 < x_2 <x_1$ has the form
\eq
\alpha_{321} \,  e^{i (k_1 x_1 + k_2 x_2 + k_3 x_3) }, \quad  \alpha_{inc}=(-P_{13})\alpha_{321},
\en
where the permutation operator $P_{13}$ exchanges the spins of the first qubit and the third one,
so that the scattering matrix of the three qubits in different regions is given by
\eq
\alpha_{out} = \check{R}_{12}(k_2,k_1)\,\check{R}_{23}(k_3,k_1)\,
\check{R}_{12}(k_3,k_2)\,(-P_{12})\,(-P_{23})\,(-P_{12})  \alpha_{inc}
\en
where $P_{13}=P_{12}\,P_{23}\,P_{12}$ is used.

In term of the notation $\check{R}(u)$, the identity operator $Id$ and the permutation operator $P$,
the two-body qubit gates $\check{R}_{12}$ and $\check{R}_{23}$ have a new form
\eq
\label{ru}
 \check{R}(u)  = \frac {-1} {i u +c} (c + i u P),\quad   \check{R}_{12}=\check{R} \otimes Id,
   \qquad \check{R}_{23}=Id \otimes \check{R},
\en
where the two-qubit gate $\check{R}(u)$ satisfies the unitarity condition,
\eq
\check{R}(u)\, \check{R}^\dag(u) = \check{R}^\dag(u)\,  \check{R}(u)  =Id, \quad  \check{R}^\dag(u)=\check{R}(-u).
\en
 Hence  the multi-qubit factorisable scattering with the delta-function interaction has been
 recognized as a quantum circuit  model consisting of the Swap gate $P$ and the two-qubit gate
 $\check{R}(u)$ (\ref{ru}).

Besides the form (\ref{ru}), the two-qubit gate $\check{R}(u)$ has the other interesting formalism
\eq
 \check{R}(\varphi)=-e^{-2 i \varphi} P_-  -  P_+, \quad P_\pm =\frac {1 \pm P} 2.
\en
with $\varphi=\arctan {\frac u c}$ and $-\frac \pi 2 < \varphi < 0$. At
$u=-c$, the two-qubit gate $\check{R}(u)$ is the $\sqrt{Swap}$ gate given by
\eq
\check{R}(-\frac \pi 4)= - \sqrt{Swap},\quad  \sqrt{Swap} = P_{+} + i P_{-}
\en
which is an entangling two-qubit gate but can not yield all
unitary $4 \times 4$ matrices by itself (even with the Swap gate) \cite{LD98}.

Furthermore,  the two-qubit gate $\check{R}(\varphi)$ has the  form of the time evolutional
operator  $U(\varphi)$ of the Heisenberg  interaction $\vec{S}_1\cdot \vec{S}_2$ between two
qubits
 \eq
 \check{R}(\varphi) = -e^{-i \frac \varphi 2} U(-2 \varphi),
  \quad U(\varphi)=e^{-i \varphi \vec{S}_1\cdot\vec{S}_2}
 \en
modulo a global phase factor with the permutation $P=2 \vec{S}_1\cdot \vec{S}_2 +\frac 1 2$
so that $\check{R}(\varphi)$ and $U(\varphi)$ are two equivalent two-qubit gates in the
quantum circuit model.

As is realized by DiVincenzo et al. \cite{DiVincenzo00},   universal quantum computation
can be set up only using the Heisenberg interaction when a logical qubit is encoded as a
two-dimensional subspace of eight-dimensional Hilbert space of three qubits.
Hence, the quantum circuit model in terms of the Swap gate $P$ and the two-qubit gate
$\check{R}(u)$ (\ref{ru}) is also able to perform universal quantum computation, if
and only if a logical qubit is chosen in a suitable way.

\section{Quantum Computing via the Yang-Baxter equation}

 A quantum circuit model in terms of both a nontrivial unitary solution $\check{R}(u)$ of the
 Yang--Baxter equation (\ref{YBE-three}) and the Swap gate $P$ is called {\em the generalized
 factorisable scattering model} in the paper. Once it is given, the problem becomes how to find out the
 underlying physical system,  which is thought of as the inverse problem of quantum computing
 via the Bethe ansatz.

In the literature, the Yang--Baxter equation has the usual formalism,
\eq
 \label{ybe}
 (\check{R}(x)\otimes 1\!\! 1_2)\,(1\!\! 1_2\otimes \check{R}(x y))\,(\check{R}(y)\otimes 1\!\! 1_2)
 =
 (1\!\! 1_2\otimes \check{R}(y))\, (\check{R}(x y)\otimes 1\!\! 1_2)\, (1\!\! 1_2\otimes \check{R}(x))
 \en
where the two-qubit gate $\check{R}(x)$  is a linear operator on the Hilbert space of two qubits, i.e.,
$\check{R}: {\cal H} \otimes {\cal H} \to {\cal H}\otimes {\cal H}$,  $1\!\! 1_2$ denotes the $2\times 2$
identity matrix, and $x, y$ are called the spectral parameter. Take $x=e^u$ and $y=e^v$, then the formalism of
the Yang--Baxter equation (\ref{YBE-three}) can be derived. As a two-qubit gate,  the solution $\check{R}(x)$
of the Yang--Baxter equation (\ref{ybe}) has to satisfy the unitary condition
\eq  \label{unitary}
\check{R}(x)\,\check{R}^\dag(x)=\check{R}^\dag(x)\,\check{R}(x)=
\rho 1\!\! 1_4 \en
with the normalization factor $\rho$.

The construction of the Hamiltonian via a nontrivial unitary solution of the Yang--Baxter equation (\ref{ybe})
is a very flexible or subtle process with physical reasoning, namely, there does not exist a universal law to
guide such a construction, partly because the same solution of the Yang--Baxter equation can be yielded
by many different kinds of physical interactions.

The two-qubit gate (\ref{ru}), as a rational solution of the Yang--Baxter equation (\ref{ybe}),
has the form
\eq
 \check{R}(\alpha) = - \frac 1 {1+\alpha} (1\!\! 1_4+\alpha P), \quad \alpha= i \frac u c
\en
which is a linear combination of the identity $1\!\! 1$ and the permutation $P$. In terms of the variable $\varphi$,
the two-qubit gate  (\ref{ru}) can be expressed as an exponential formalism of the permutation $P$ given by
 \eq
 \label{rv}
  \check{R}(\varphi)=e^{i(\pi-\varphi)} e^{i \varphi P}, \quad  \tan\varphi =  \frac u c,
 \en
hence the Hamiltonian for the generalized factorisable scattering model can be chosen as the Heisenberg
interaction $\vec{S}_1\cdot\vec{S}_2$ between two qubits. On the other hand, it is easy to examine the
delta-function interaction (\ref{delta}) as a suitable Hamiltonian underlying the two-qubit gate (\ref{ru}), but it
is not so explicit to realize the delta-function interaction (\ref{delta}) from the formalism of the two-qubit
gate (\ref{ru}).

In  \cite{ZKG05},  an approach for the construction of the Hamiltonian is presented via a two-qubit solution
$\check{R}(x)$ of the Yang--Baxter equation (\ref{ybe}). With an initial state $\psi$, the evolution
state $\psi(x)$ determined by $\check{R}(x)$ is given by
 $\psi(x)=\rho^{-\frac 1 2}\check{R}(x)\psi$. The  Shr{\" o}dinger equation has the form
 \eq i\,\frac {\partial \psi(x)} {\partial x}=H(x)\psi(x) \en  with the Hamiltonian $H(x)$ given by
  \eq \label{hx}   H(x)=i\,\frac {\partial }{\partial x}(\rho^{-\frac 1 2}\check{R})
  \rho^{-\frac 1 2}\check{R}^{-1}(x) \en
which is time-dependent of $x$.

In terms of an appropriate parameter instead of $x$, the Hamiltonian $H(x)$ (\ref{hamiltonian}) may be transformed
to a time-independent formalism. For example, the Hamiltonian $H(u)$ (\ref{hamiltonian})  via the two-qubit
gate (\ref{ru}) is obtained to be time-dependent of the parameter $u$ after some algebra using the formula
(\ref{hx}), but the Hamiltonian via the the formalism (\ref{rv}) of this gate is found to be time-independent
Heisenberg interaction.

Consider the other  solution of the Yang--Baxter equation (\ref{ybe}),
\eq  \label{belltype1}
\check{R}(x)= \left(\begin{array}{cccc}
1+x & 0 & 0 & 1-x \\
0 & 1+x & -(1-x) & 0 \\
0 & 1-x & 1+x & 0 \\
-(1-x) & 0 & 0 & 1+x \end{array}\right), \en
where the unitarity condition (\ref{unitary}) requires $x$ real with the normalization factor $\rho=2(1+x^2)$.
In terms of the variable $\theta$ defined by \eq \label{belltypetran} \cos\theta=\frac 1
{\sqrt{1+x^2}}, \qquad
 \sin\theta=\frac x {\sqrt{1+x^2}}, \qquad
 \en
the two-qubit gate $\check{R}(x)$ (\ref{belltype1}) has the  formalism of $\theta$,
and the Hamiltonian $H(\theta)$ (\ref{hamiltonian}) is calculated to be \eq \label{hamiltonian}
 H(\theta)=\frac i 2 \frac {\partial  } {\partial \theta}(\rho^{-\frac 1 2}\check
 R) \rho^{-\frac 1 2}
 \check R^\dag=\frac 1 2\sigma_x\otimes \sigma_y \en
with the conventional form of the Pauli matrices. The time evolutional operator has the form
$U(\theta)=e^{-i H\theta}$ as well as the two-qubit gate (\ref{belltype1}) is given by
 \eq \check{R}(\theta)=\cos(\frac \pi 4-\theta)+ 2\,i\,
 \sin(\frac \pi 4-\theta)\,H=e^{i (\frac \pi
 2-2\,\theta) H}. \en
At $\theta=0$, the two-qubit gate $\check R(0)$ given by
 \eq
  \check R(0)
 = \left( \begin{array}{cccc}
1/\sqrt{2} & 0 & 0  & 1/\sqrt{2}\\
0  & 1/\sqrt{2} & -1/\sqrt{2} & 0\\
0 & 1/\sqrt{2} & 1/\sqrt{2} & 0 \\
-1/\sqrt{2} & 0 & 0 & 1/\sqrt{2}\\
\end{array} \right)
 \en
is a unitary basis  transformation matrix from the product base to the Bell states.

Besides the $\check{R}(x)$ matrix (\ref{belltype1}), in \cite{ZKG05},  the following type of
nontrivial unitary solutions $\check{R}(x)$ of the Yang--Baxter equation (\ref{ybe}),
 \eq \label{matchgate}  \check{R}(x) =\left(\begin{array}{cccc}
  \omega_1(x) & 0 & 0 & \omega_7(x) \\
  0 & \omega_5(x) & \omega_3(x) & 0 \\
  0 & \omega_4(x) & \omega_6(x) & 0 \\
  \omega_8(x) & 0 & 0 & \omega_2(x)
  \end{array}\right),\en
have been recognized as two-qubit  gates modulo a phase factor as well as the associated
time-dependent Hamiltonians $H(x)$ (\ref{hx}) also respectively calculated. The two-qubit
gate $\check{R}(x)$ (\ref{matchgate})  is determined by two submatrices $U_1$ and $U_2$,
\eq   U_1(x) =\left(\begin{array}{cc}
  \omega_1(x) &   \omega_7(x) \\
  \omega_8(x) &  \omega_2(x)
  \end{array}\right),\quad  U_2(x) =\left(\begin{array}{cc}
  \omega_5(x) & \omega_3(x) \\
  \omega_4(x) & \omega_6(x)
  \end{array}\right),   \en
which are in the unitary group $U(2)$.

According to Terhal and DiVincenzo's understanding \cite{TD02} of Valient's work on matchgates,  when
the above unitary matrices $U_1$ and $U_2$ are elements of the special unitary group $SU(2)$, quantum
computations with the $\check{R}(x)$ gate (\ref{matchgate}) acting on nearest-neighbor qubits can be 
efficiently simulated on a classical computer, whereas the $\check{R}(x)$ gate (\ref{matchgate}) combined 
with the Swap gate acting on farther-neighbor qubits may be capable of performing universal quantum 
computation. Obviously, the two-qubit gate (\ref{belltype1}) has its two submatrices $U_1$ and $U_2$ in 
the $SU(2)$ group.

In terms of $U_1$ and $U_2$ in $SU(2)$,  the other two-qubit gate with six nonvanishing entries given by
\eq  \check{R}(\theta) =\left(\begin{array}{cccc}
  \sinh(\gamma-i\theta)) & 0 & 0 & 0 \\
  0 & e^{i\theta} \sinh\gamma  & -i \sin\theta & 0 \\
  0 & -i\sin\theta & e^{-i\theta}\sinh\gamma & 0 \\
  0 & 0 & 0 & \sinh(\gamma+i\theta)
\end{array}\right),\en
with the normalization factor $\rho=\sinh^2\gamma +\sin^2\theta$, can be found in \cite{ZKG05} as a
unitary solution of the Yang--Baxter equation (\ref{YBE-three}) with  the spectral parameter $\theta$ and
real parameter $\gamma$, but which gives rise to a time-dependent Hamiltonian (\ref{hamiltonian}).

However, the two-qubit gate (\ref{rv}) associated with the delta-function interaction (\ref{delta}) or
Heisenberg interaction consists of
the following submatrices $U_1$ and $U_2$ given by
\eq   U_1 =-\left(\begin{array}{cc}
  1 &   0 \\
  0 &  1
  \end{array}\right),\quad  U_2 =-e^{-i\varphi}  \left(\begin{array}{cc}
  \cos\varphi & i\sin\varphi \\
  i\sin\varphi & \cos\varphi
  \end{array}\right),   \en
where $U_1$ is in $SU(2)$ but $U_2$ is in $U(2)$ because of the global phase. With the encoded logical qubit,
it has been numerically verified \cite{DiVincenzo00} that the two-qubit gate (\ref{rv}) itself can achieve
universal quantum  computation by acting on nearest-neighbor qubits.

\section{Concluding remark}

Feynman is well known in the community of quantum information science due to  his
pioneering work on both universal quantum simulation in 1982 and quantum circuit model in 1986.  Shortly before
he passed away in early 1988, very unexpectedly, Feynman wrote: ``I got really fascinated by these (1+1)-dimensional
models that are solved by the Bethe ansatz", see Batchelor's feature article \cite{Batchelor07} on the research history
of the Bethe ansatz.

A definition of {\em integrable quantum computation} in both physics and mathematics has been proposed in
this article, in order to remove potential conceptual confusions between the papers \cite{Zhang11}
and \cite{ZKG05} as well as declare {\em integrable quantum computation} as an independent research subject in
quantum information and computation. Note that {\em integrable quantum computation} has been also argued from
the viewpoint of the Hamiltonian formalism of quantum error correction codes \cite{Zhang08}.

The asymptotic condition $\check{R}(x=0)$ of the solution of  the Yang--Baxter equation (\ref{ybe}) satisfies
the braid group relation and hence $\check{R}(x=0)$ can be viewed as a unitary braiding gate, which suggests
similarities and comparisons between {\em integrable quantum computation} and quantum computing via unitary
braid representations, see \cite{ZKG05, Zhang08} for the detail and  \cite{Zhang11} for comments.

\section*{Acknowledgements}

The author wishes to thank Professor Lu Yu and Institute of Physics,
Chinese Academy of Sciences, for their hospitality and support during
the visit in which this work was done and the part of the previous work
\cite{Zhang11} had been done.

 \end{document}